\documentclass[12pt]{article}
\usepackage{a4}
\usepackage{amsmath}
\usepackage{graphics}
\usepackage{graphicx}
\usepackage{latexsym}
\usepackage{amsfonts}
\usepackage{graphics}
\usepackage{epsfig}
\usepackage{epic}

\author{T.~Leonhardt, R.~Manvelyan and W.~R\"uhl}
\title{The group approach to AdS space propagators}

\begin{document}
\renewcommand{\thefootnote}{\fnsymbol{footnote}}
\thispagestyle{empty}

\noindent hep-th/   \hfill  February  2004 \\

\noindent \vskip3.3cm
\begin{center}

{\Large\bf Coupling of Higher Spin Gauge Fields to a Scalar Field
in $AdS_{d+1}$ and their Holographic Images in the $d$-Dimensional
Sigma Model}
\bigskip\bigskip\bigskip

{\large Thorsten Leonhardt, Ruben Manvelyan\footnote{On leave from
Yerevan Physics Institute, e-mail: manvel@moon.yerphi.am} and
Werner R\"uhl}
\medskip

{\small\it Department of Physics\\ Erwin Schr\"odinger Stra\ss e \\
University of Kaiserslautern, Postfach 3049}\\ {\small\it 67653
Kaiserslautern, Germany}\\
\medskip
{\small\tt tleon,manvel,ruehl@physik.uni-kl.de}
\end{center}

\bigskip 
\begin{center}
{\sc Abstract}
\end{center}
The three-point functions of two scalar fields $\sigma$ and the
higher spin field $h^{(\ell)}$ of $HS(4)$ on the one side and of
their proposed holographic images $\alpha$ and
$\mathcal{J^{(\ell)}}$ of the minimal conformal $O(N)$ sigma model
of dimension three on the other side are evaluated at leading
perturbative order and compared in order to fix the coupling
constant of $HS(4)$. This necessitates a careful analysis of the
local current $\Psi^{(\ell)}$ to which $h^{(\ell)}$ couples in
$HS(4)$ and which is bilinear in $\sigma$.

\noindent

\newpage
\renewcommand{\thefootnote}{\arabic{footnote}}\setcounter{footnote}{0}
\section{Introduction}
 \quad In any known model the AdS/CFT correspondence
is an unproven hypothesis still. If such model is derived from
string theory as the standard case of $AdS_{5}$ supergravity and
 $SYM_{4}(\mathcal{N}$=4), supersymmetry permits geometric
arguments based on representation theory that support AdS/CFT
correspondence and these arguments look quite convincing indeed.

But in models of the type of higher spin gauge fields (HS(d+1))
there is no supersymmetry a priori and the correspondence can be
proved only by dynamical calculations both in $AdS_{d+1}$ and
$CFT_{d}$ cases. Since in these models perturbative expansions
with small coupling constants are mapped on each other, such
calculations are technically feasible and the holographic mapping
is order by order.We shall start such calculation for $HS(4)$ and
the 3-dimensional conformal $O(N)$ sigma model now.

We concentrate on three-point function of two scalar and one
higher spin field
\begin{equation}\label{fields}
\begin{array}{|c|c|c|}
\hline
   & AdS_{4} & CFT_{3} \\
\hline
 \textnormal{Scalar} & \sigma(z)& \alpha(x) \\
\hline
 \textnormal{HSF} & h^{(\ell)}(z) & \mathcal{J}^{(\ell)}(x) \\
\hline
\end{array}    \nonumber
\end{equation}
where $\alpha(x)$ is the ``auxiliary'' or ``Lagrange multiplier''
field and $\mathcal{J^{(\ell)}}(x)$ an almost conserved current,
which is a traceless symmetric tensor. In the sigma model case the
coupling constant is $O(\frac{1}{\sqrt{N}})$. In the higher spin
field theory the coupling constant for $\sigma\sigma h^{(\ell)}$
interaction is $g_{\ell}$, so that we expect
\begin{equation}\label{gl}
    g_{\ell}=C^{(\ell)} \frac{1}{\sqrt{N}} \;.
\end{equation}
We determine $C^{(\ell)}$ first in an ad hoc wave function
normalization such that
\begin{eqnarray}
  &&\langle \alpha(x)\, \alpha(0)\rangle_{CFT} =\left(x^{2}\right)^{-\beta}\label{2palpha}\\
   &&\langle \sigma(z_{1})\,\sigma(z_{2})\rangle_{AdS}=
   \left(2\zeta\right)^{-\beta}
   F\left[ {\frac{1}{2}\beta,\frac{1}{2}(\beta+1)\atop \beta-\mu+1
   } ;\; \zeta^{-2} \right]\label{2psigma} \\
 &&\zeta=\frac{(z_{1}^{0})^{2}+z_{2}^{0})^{2}+
 (\vec{z}_{1}-\vec{z}_{2})^{2}}{2z^{0}_{1}z^{0}_{2}}\quad, \quad
 \mu=\frac{1}{2}d\;,
\end{eqnarray}
so that (\ref{2palpha}) is obtained from (\ref{2psigma}) by a
``simple'' boundary limit
\begin{equation}\label{limit}
    \lim_{z^{0}_{1}\rightarrow 0}\lim_{z^{0}_{2}\rightarrow
    0}\left(z^{0}_{1}z^{0}_{2}\right)^{-\beta}\langle
    \sigma(z_{1}) \,\sigma({z_{2}})\rangle_{AdS}=\langle
    \alpha(\vec{z}_{1})\,
    \alpha(\vec{z}_{2} )\rangle_{CFT}\;.
\end{equation}
The higher spin fields are assumed to be normalized in the same
fashion. At the end we renormalize the higher spin field such that
$C^{(\ell)}$ is replaced by one.

We shall treat two versions of the minimal $O(N)$ sigma model. In
the ``free'' case we have as a scalar field
\begin{equation}\label{alphafree}
\alpha_{f}(x)=\frac{1}{\sqrt{2N}}\phi_{i}(x)\;\phi_{i}(x)\;,
\end{equation}
where $\phi_{i}(x), i=1,2,\dots N$  is the $O(N)$ vector and
space-time scalar field normalized so that
\begin{equation}\label{2pphi}
    \langle\phi_{i}(x)\,\phi_{j}(x)
    \rangle_{CFT}=\left(x^{2}\right)^{-\delta}\delta_{ij}\quad,\quad
    \delta=\mu-1
\end{equation}
and (\ref{2palpha}) follows from (\ref{2pphi}) and
(\ref{alphafree}) with
\begin{equation}\label{beta}
    \beta_{f}=2(\mu-1)=d-2 \;.
\end{equation}

In the ``interacting'' sigma model we have an interaction
\begin{equation}\label{sigmaint}
    \textbf{z}^{1/2}\int dx \, \phi_{i}(x)\;\phi_{i}(x)\,\alpha(x)
\end{equation}
and the interaction constant $\textbf{z}$ is expanded
\begin{equation}\label{expansion}
\textbf{z}=\sum^{\infty}_{k=1}\frac{\textbf{z}_{k}}{N^{k}} \;.
\end{equation}
The ``free'' theory is unstable and by renormalization flow
approaches the stable ``interacting'' theory. The conformal scalar
field $\sigma(z)$ on $AdS_{d+1}$  is massive (tachyonic due to
conformal coupling with the AdS metric) and has two boundary
values from the two roots of the dimension formula
\begin{equation}\label{dimform}
    \Delta=\mu\pm \left(\mu^{2}+m^{2}\right)^{\frac{1}{2}} \;,
\end{equation}
where for $d=3$
\begin{equation}\label{m2}
    m^{2}=\left\{{\hspace{-2.8cm}-2 \quad\textnormal{in the free case}\atop -2
    +O(\frac{1}{N}) \quad\textnormal{in the interacting case}}\right.
\end{equation}
so that
\begin{equation}\label{delta}
   \Delta(d=3)=\left\{{\hspace{-1.3cm}\beta_{f}=1\quad\textnormal{from (\ref{beta})}\atop \beta=2
    +O(\frac{1}{N}) \quad\textnormal{from
    (\ref{sigmaint})}\;.}\right.
\end{equation}

\section{Currents coupled to (conformal)\\ higher spin fields in AdS}

We assume that the interaction of a spin $\ell $ gauge field
$h^{(\ell)}$ and two scalar fields $\sigma(z)$ is local and
mediated by a current $\Psi^{(\ell)}$
\begin{equation}\label{int}
    \int\frac{dz}{(z^{0})^{d+1}}Tr\left\{\Psi^{(\ell)}(z)\,h^{(\ell)}(z)\right\}\;.
\end{equation}
$\Psi^{(\ell)}$ and $h^{(\ell)}$ are symmetric tensors of rank
$\ell$. If we postulate that the covariant divergence of
$\Psi^{(\ell)}$ is a trace term, the interaction is gauge
invariant. Namely a gauge transformation of $h^{(\ell)}$, being of
the form (classical)
\begin{equation}\label{gt}
    h^{(\ell)} \rightarrow h^{(\ell)} + \nabla\Lambda^{(\ell
    -1)}\;,
\end{equation}
where $\Lambda^{(\ell -1)}$ is a symmetric traceless tensor and
$\nabla\Lambda^{(\ell -1)}$ is symmetrized, leads to the zero
gauge variation of (\ref{int})
\begin{equation}\label{intgauge}
    Tr\left\{\nabla \Psi^{(\ell)}\Lambda^{(\ell -1)}\right\}=0 \;.
\end{equation}

This consideration is in agreement with the so called "Fronsdal"
 theory \cite{Fronsdal} of higher spin with double-traceless
gauge fields and currents. Truncation of this higher spin theory
to the \emph{conformal higher spin theory} can be observed if we
consider the corresponding double-traceless current and gauge
field as a sum of two traceless objects, namely
\begin{equation}\label{double}
\Psi^{(\ell)}=J^{(\ell)}+g^{(2)}\psi^{(\ell -2)}\;,
\end{equation}
where $J^{(\ell)}$ and $\psi^{(\ell -2)}$ are now the
\emph{traceless} tensors, $g^{(2)}$ is the $D=d+1$ dimensional AdS
metric and symmetrization is assumed. Following \cite{Segal} we
call $\psi^{(\ell -2)}$ the compensator field. It is easy to see
that $\psi^{(\ell -2)}$ plays the role of the traceless trace of
the double-traceless current $\Psi^{(\ell)}$ and has to decouple
in the conformal limit of higher spin theory. In other words we
will assume that at the $d$ dimensional boundary $M_{d}=\partial
AdS_{D}$ , $\Psi^{\ell}$ behaves as a conformal tensor field. Now
we will consider the general structure of conformal higher spin
currents in the $AdS_{D}$ space constructed from the conformally
coupled scalar field $\sigma(z)$ with the corresponding on-shell
condition
\begin{equation}\label{on-shell}
\Box \sigma(z)=\nabla \cdot \nabla
\sigma(z)=\frac{D(D-2)}{4L^{2}}\sigma(z)\;.
\end{equation}

The tachyonic mass here (we use in this section the mainly minus
signature of the AdS metric \footnote{We will use AdS conformal
flat metric, curvature and covariant derivatives comutation rules
of the type
\begin{eqnarray}\label{ads}
&&ds^{2}=g_{\mu\nu}dz^{\mu}dz^{\nu}=\frac{L^{2}}{(z^{0})^{2}}\eta_{\mu\nu}dz^{\mu}dz^{\nu},\quad
\eta_{z^{0}z^{0}}=-1, \sqrt{-g}=\frac{1}{(z^{0})^{d+1}}\;,\nonumber\\
&&\left[\nabla_{\mu},\,\nabla_{\nu}\right]V^{\rho}_{\lambda} =
R^{\quad\,\,\rho}_{\mu\nu \sigma}V^{\sigma}_{\lambda}
  -R^{\quad\,\,\sigma}_{\mu\nu \lambda}V^{\rho}_{\sigma}\;,\\
 &&R^{\quad\,\,\rho}_{\mu\nu \lambda}=
-\frac{1}{(z^{0})^{2}}\left(\eta_{\mu\lambda}\delta^{\rho}_{\nu} -
\eta_{\nu\lambda}\delta^{\rho}_{\mu}\right)=-\frac{1}{L^{2}}
\left(g_{\mu\lambda}\delta^{\rho}_{\nu}
- g_{\nu\lambda}\delta^{\rho}_{\mu}\right) \;,\nonumber\\
 &&R_{\mu\nu}=-\frac{D-1}{(z^{0})^{2}}\eta_{\mu\nu}=-\frac{D-1}{L^{2}}g_{\mu\nu}\quad ,
\quad R=-\frac{D(D-1)}{L^{2}}\;.\nonumber
\end{eqnarray}}) arises as a result of conformal coupling of the
conformal scalar $\sigma(z)$ with the AdS curvature $S=\int
d^{D}z\sqrt{-g}\frac{1}{2}\left(g^{\mu\nu}\partial_{\mu}\sigma
\partial_{\nu}\sigma-\frac{D-2}{4(D-1)}R\sigma^{2}\right)$.
For the investigation of the conservation and tracelessness
conditions for general spin $\ell $ symmetric conformal current
$J^{(\ell)}_{\mu_{1}\mu_{2}\dots\mu_{\ell}}$ we contract it with
the $\ell$-fold tensor product of a vector $a^{\mu}$ and make the
ansatz including a first curvature correction in contrast to the
free flat case \cite{Anselmi}
\begin{eqnarray}
  &&J^{(\ell)}(z;a) = \frac{1}{2}\sum^{\ell}_{p=0}A_{p}\left(a\nabla\right)^{\ell
  -p}\sigma(z)\left(a\nabla\right)^{p}\sigma(z) \nonumber\\
  &&+ \frac{a^{2}}{2}\sum^{\ell -1}_{p=1}B_{p}\left(a\nabla\right)^{\ell
  -p-1}\nabla_{\mu}\sigma(z)\left(a\nabla\right)^{p-1}\nabla^{\mu}\sigma(z)\label{ansatz}\\
  &&+\frac{a^{2}}{2L^{2}}\sum^{\ell -1}_{p=1}C_{p}\left(a\nabla\right)^{\ell
  -p-1}\sigma(z)\left(a\nabla\right)^{p-1}\sigma(z) + O(a^{4}) +
  O(\frac{1}{L^{4}})\;,\nonumber
\end{eqnarray}
where $A_{p}=A_{\ell-p}, B_{p}=B_{\ell-p}, C_{p}=C_{\ell-p}$ and
$A_{0}=1$. Now we  try to define the set of unknown constants
$A_{p}, B_{p}$ and $C_{p}$ using the current conservation
condition
\begin{equation}\label{curcons}
\nabla \cdot
\partial_{a}J^{(\ell)}(z;a)=\nabla^{\mu}\frac{\partial}{\partial
a^{\mu}}J^{(\ell)}(z;a)=0
\end{equation}
and the tracelessness condition connected with the conformal
nature of our scalar field $\sigma(z)$
\begin{equation}\label{trace}
    \Box_{a}J^{(\ell)}(z;a)=\frac{\partial^{2}}
    {\partial a_{\mu}\partial a^{\mu}}J^{(\ell)}(z;a)=0 \;.
\end{equation}
Using the following basic relations
\begin{eqnarray}
  &&\left[\nabla_{\mu}, \left(a\nabla\right)^{p}\right]\sigma =
  \frac{p(p-1)}{2L^{2}}\left(a_{\mu}\left(a\nabla\right)^{p-1}
  \sigma - a^{2}\left(a\nabla\right)^{p-2}\nabla_{\mu}\sigma\right)\;, \\
 &&\left[\nabla_{\mu}, \left(a\nabla\right)^{p}\right]\nabla_{\nu}\sigma =
  \frac{p(p-1)}{2L^{2}}\left(a_{\mu}\left(a\nabla\right)^{p-1}\nabla_{\nu}
  \sigma -
  a^{2}\left(a\nabla\right)^{p-2}\nabla_{\mu}\nabla_{\nu}\sigma\right)\nonumber\\
  &&\,\,\quad\qquad\qquad\qquad+\frac{p}{L^{2}}\left(g_{\mu\nu}\left(a\nabla\right)^{p}\sigma -
  a_{\nu}\left(a\nabla^{p-1}\nabla_{\mu}\sigma\right)\right)\;,\\
  &&\frac{\partial}{\partial a^{\mu}}\left(a\nabla\right)^{p}\sigma=
  p\left(a\nabla\right)^{p-1}\nabla_{\mu}\sigma\nonumber\\&&\quad\qquad\qquad\qquad + \frac{p(p-1)(p-2)}{6L^{2}}\left(a_{\mu}\left(a\nabla\right)^{p-2}
  \sigma - a^{2}\left(a\nabla\right)^{p-3}\nabla_{\mu}\sigma\right)\;,\qquad\\
  && \nabla \cdot \frac{\partial}{\partial a}\left(a\nabla\right)^{p}\sigma=
  \frac{1}{L^{2}}\left[\frac{1}{4}pD(D-2)\right.\nonumber\\&&\left.
  \quad\qquad\qquad\qquad +p(p-1)\left(D +
  \frac{2}{3}p-\frac{7}{3}\right) \right]\left(a\nabla\right)^{p-1}\sigma + O(\frac{1}{L^{4}})\;,\\
  &&\Box_{a}\left(a\nabla\right)^{p}\sigma =
  \frac{1}{L^{2}}\left[\frac{1}{4}p(p-1)D(D-2)\right.\nonumber\\
  &&\left.\quad\qquad\qquad\qquad +\frac{1}{3}p(p-1)(p-2)(p+2D-5)\right]
  \left(a\nabla\right)^{p-2}\sigma + O(\frac{1}{L^{4}})\qquad
  \end{eqnarray}
  we can derive recursion relations for $A_{p}$, $B_{p}$ and
  $C_{p}$ coming from conservation condition (\ref{curcons})
  \begin{eqnarray}
    &&pA_{p} + (\ell -p +1)A_{p-1} + 2B_{p} + 2B_{p-1}=0 , \label{flat}\\
    &&s_{3}(p)A_{p+1} + s_{2}(p,\ell , D)A_{p} +s_{2}(\ell -p+1,\ell , D)A_{p-1} \nonumber\\
    &&+
    s_{3}(\ell -p +1)A_{p-2} +2C_{p} + 2C_{p-1}=0  , \label{L2}\\
    &&s_{2}(p,\ell , D)= \frac{1}{4}pD(D-2)+p(p-1)(D
    +\frac{1}{2}\ell+\frac{1}{6}p-\frac{7}{3}) ,\\
    &&s_{3}(p)=\frac{1}{6}(p+1)p(p-1) \;.
   \end{eqnarray}
The relation (\ref{flat}) relates $A_{p}$ and $B_{p}$ recursively
as in the flat case \cite{Anselmi}. The next relation (\ref{L2})
arises from the $\frac{1}{L^{2}}$ correction and relates
recursively $C_{p}$ and $A_{p}$ coefficients from our ansatz
(\ref{ansatz}). From the other side the tracelessness condition
(\ref{trace}) gives us two further  relations between these
coefficients
\begin{eqnarray}
  &&B_{p}=-\frac{p(\ell -p)}{(D+2\ell-4)}A_{p} ,\label{flatT} \\
  &&C_{p}=\frac{-1}{2(D+2\ell -4)}\left[s_{t}(p+1,\ell,D)A_{p+1}
  +s_{t}(\ell -p+1,\ell, D)A_{p-1}\right], \quad\label{L2T}\\
  &&s_{t}(p,\ell,D)=\frac{1}{4}p(p-1)D(D-2) +
  \frac{1}{3}p(p-1)(p-2)(\ell +2D-5) \;.
\end{eqnarray}
Again the relation (\ref{flatT}) is the same as in the flat case
and leads the Eq. (\ref{flat}) to the recursion
\begin{eqnarray}
  A_{p} &=& -s_{1}(p,\ell ,D)A_{p-1},\label{rec1} \\
  s_{1}(p,\ell ,D)&=&\frac{(\ell -p+1)(2\ell -2p+D-2)}{p(D+2p-4)}
  \;.
\end{eqnarray}
From this we can obtain the same  solution for the $A_{p}$
coefficients \cite{Anselmi} as in the flat case
\begin{equation}\label{so1}
    A_{p}=(-1)^{p}\frac{\binom{\ell}{p}\binom{\ell +D-4}{p+\frac{D}{2}-2}}
    {\binom{\ell +D-4}{\frac{D}{2}-2}} \;.
\end{equation}
For the important  case $D=4$  this formula  simplifies to
\begin{equation}\label{4d}
 A_{p}=(-1)^{p}\binom{\ell}{p}^{2} \;.
\end{equation}
It means that if our ansatz (\ref{ansatz}) and our consideration
for the $\frac{1}{L^{2}}$ correction are right,  the recursion
relation for the $A_{p}$ coefficients obtained by substituting the
$C_{p}$ coefficients in (\ref{L2}) by those of the
$\frac{1}{L^{2}}$ tracelessness condition (\ref{L2T}) must be
consistent with (\ref{rec1}). Indeed using (\ref{L2T}) and
(\ref{rec1}) we can rewrite the relation (\ref{L2}) in the form
\begin{eqnarray}
  &&(\ref{L2})= A_{p}s_{f}(p,\ell ,D) +A_{p-1}s_{f}(\ell -p+1,\ell,D)=0 \label{L21}\\
  && s_{f}(p,\ell ,D) = \left[s_{2}(p,\ell ,D) -
  \frac{s_{t}(p,\ell ,D)}{D+2\ell -4}\right.\nonumber\\&&\left.-s_{1}(p+1,\ell ,D)\left(s_{3}(p)-
  \frac{s_{t}(p+1,\ell ,D)}{D+2\ell -4}\right)\right]\\
  &&=\frac{(\ell +D-3)(2\ell
  +D-2)p(D+2p-4)}{4(D+2\ell-4)}\;.\nonumber
\end{eqnarray}
It is easy to see that the relation (\ref{L21}) coincides with
(\ref{rec1}) because
\begin{equation}\label{cf}
    \frac{s_{f}(p,\ell ,D)}{s_{f}(\ell
    -p+1,\ell,D)}=s_{1}(p,\ell,d) \;.
\end{equation}

So we obtain a result that the structure of the conformal higher
spin currents constructed from the conformal coupled scalar field
in the fixed AdS background remains the same as in the free flat
space case. We prove that our ansatz with $\frac{1}{L^{2}}$
correction connected with the difference between the traces in
flat and AdS case does not violate the conservation condition
(recursion relation (\ref{rec1})) for the coefficients $A_{p}$ if
they obey the tracelessness condition (\ref{L2T}) for the
currents. It means that the traceless conserved higher spin
current constructed from conformal scalar field in AdS can be
obtained from the flat space expression replacing usual
derivatives with covariant ones and adding corresponding curvature
corrections to the expression for the traces. For completeness we
present in the Appendix an explicit derivation of the conformal
conserved current in the case $\ell =4, D=4$ in all orders of
$\frac{1}{L^{2}}$.

This phenomenon we can explain now in the following way: The
conformal group for $D$- dimensional flat(with $SO(D-1,1)$
isometry) and AdS space (with $SO(D-1,2)$ isometry) is the same
-$SO(D,2)$ \footnote{Note that this is about conformal group of
AdS space-not boundary}. So we can say that the conformal
primaries or the traceless conserved currents are the same due to
the $\frac{1}{L^{2k}}$ corrections. But these originate from the
curvature corrections to the flat space equation of motion and
noncommutativeness of the covariant derivatives. Then because all
currents are \emph{traceless}  we get the cancellation of all
$\frac{1}{L^{2k}}$ accompanying terms coming from these two
sources of deformation of the flat case relations  in the
conservation condition (\ref{curcons}).

Now we will fix the coefficients $A_{p}$ from the $CFT$
consideration. We assume that on the boundary $\partial
AdS_{d+1}$, $\Psi^{(\ell)}$ behaves as a conformal tensor field
(the trace is decoupled). Moreover this conformal tensor must be
local bilinear in $\alpha(x)$ of rank $\ell$ and of dimension
\begin{equation}\label{dimalpha}
    2\beta+\ell + O(\frac{1}{N}) \;.
\end{equation}
For this purpose we evaluate the 3-point function
\begin{equation}\label{cft3p}
\langle
\alpha(x_{1})\,\alpha(x_{2})\frac{1}{2}\sum^{\ell}_{p=0}A_{p}\langle
a\cdot\partial\rangle^{p}\alpha(x_{3})\,\langle
a\cdot\partial\rangle^{\ell-p}\alpha(x_{3})\rangle_{CFT_{3}} \;,
\end{equation}
where $\langle a\cdot\partial\rangle=a^{i}\partial_{i} , \,\,
i=1,2,3$.

From the propagator (\ref{2palpha}) for $\alpha(x)$ we obtain for
(\ref{cft3p})
\begin{eqnarray}\label{3pfin}
   && 2^{\ell}\sum_{p=0}^{\ell}A_{p}(\beta)_{p}(\beta)_{\ell-p}
   \frac{\left(x^{2}_{13}x^{2}_{23}\right)^{-\beta}}{x^{2p}_{13}x^{2(\ell-p)}_{23}}
   \langle a\cdot x_{13}\rangle^{p}
   \langle a\cdot x_{23}\rangle^{(\ell-p)}  \\
    &&\hspace{8cm}+ \quad \textnormal{trace terms} \;,\nonumber
    \end{eqnarray}
where we define the Pochhammer symbols $(z)_n =
\frac{\Gamma(z+n)}{\Gamma(z)}$.

As a 3-point function of a conformal tensor is unique up to
normalization
\begin{eqnarray}
   &&
   \mathcal{C}\left(x^{2}_{13}x^{2}_{23}\right)^{-\beta}\left\{\langle a\cdot
   \xi\rangle^{\ell}+ \quad \textnormal{trace terms}\right\} ,\label{3pgen}\\
    && \xi^{i}=\frac{x_{13}^{i}}{x_{13}^{2}}-\frac{x^{i}_{23}}{x^{2}_{23}}\;,\label{xi}
    \end{eqnarray}
it follows
\begin{equation}\label{Acft}
    A_{p}=\frac{\mathcal{C}(-1)^{p}\binom{\ell}{p}}{2^{\ell}(\beta)_{p}(\beta)_{\ell-p}}\;.
\end{equation}
This expression, for $\beta=1$, is in agreement with the previous
one (\ref{4d}) obtained from $AdS_{4}$ consideration, if we will
normalize in (\ref{3pgen}) $\mathcal{C}=2^{\ell}\ell!$.

For $\beta=2$ we have to change the constraints imposed in
(\ref{ansatz}) to allow compatibility of (\ref{Acft}) with
(\ref{ansatz}). We propose to give up the condition of
tracelessness so that the coefficients in (\ref{ansatz}) are
determined from (\ref{flat}), (\ref{L2}) and (\ref{Acft}). This
implies a coupling of the first trace compensator field of
$h^{(\ell)}$ to $\Psi^{(\ell)}$.

\section{The 3-point function $\sigma\sigma h^{(\ell)}$ in $AdS_{d+1}$ to first order}
\quad Based on the interaction (\ref{int}) with the general form
of ansatz (\ref{ansatz}) we calculate the AdS 3-point function
\begin{equation}\label{3pads}
    g^{-1}_{\ell}\langle\sigma(z_{1})\,\sigma(z_{2})\,
    h^{(\ell)}_{\mu_{1}\dots\mu_{\ell}}
    \rangle\left(b_{\mu_{i}}\right)^{\ell}_{\otimes} \;.
\end{equation}

We will use from  this section the notation of Euclidian AdS with
the Euclidian scalar product $\langle\dots\rangle$ both in the
boundary and in the bulk space.

The bulk-to boundary propagator of a scalar field of dimension
$\beta$ is
\begin{equation}\label{balkboundary}
K_{\beta}(w,\vec{x})|_{\vec{x}=0}=\left(\frac{w_{0}}{w^{2}_{0}+\vec{w}^{2}}\right)^{\beta}
\end{equation}
and of a tensor field which is traceless symmetric of dimension
$\lambda$
\begin{equation}\label{bbl}
K_{\lambda}^{(\ell)}(w,\vec{x})|_{\vec{x}=0}=\frac{w_{0}^{\lambda-\ell}}
{\left(w^{2}_{0}+\vec{w}^{2}\right)^{\lambda}} \left[\langle
a,\,R(w)\vec{b}\rangle^{\ell} - \textnormal{traces}\right]\;,
\end{equation}
where
\begin{eqnarray}
   && R_{\mu\nu}(w)=\delta_{\mu\nu}-2\frac{w_{\mu}w_{\nu}}{w^{2}_{0}+\vec{w}^{2}} ,\label{R} \\
    && \sum_{\nu=0}^{d}R_{\mu\nu}(w)R_{\nu\lambda}(w)=\delta_{\mu\lambda} \;.\label{norm}
   \end{eqnarray}
This kernel $R_{\mu\nu}$ is connected with the Jacobian of the
inversion
\begin{equation}\label{inv}
    z_{\mu}\rightarrow
    z'_{\mu}=\frac{z_{\mu}}{z^{2}_{0}+\vec{z}^{2}} \;.
\end{equation}
So that
\begin{eqnarray}
   &&\frac{\partial w'_{\mu}}{\partial w_{\nu}}=\frac{1}{w^{2}_{0}+\vec{w}^{2}}R_{\mu\nu}(w) , \\
    && R_{\mu\nu}(w)=R_{\mu\nu}(w')\;.
   \end{eqnarray}
To first order in the interaction we obtain (\ref{3pads}) with all
arguments $z_{1,2,3}$ taken to the boundary
\begin{eqnarray}\label{aaj2}
&&\frac{1}{2}\sum_{n=0}^{\ell}A_{n}\int
\frac{dw}{w^{d+1}_{0}}\left(\nabla_{\mu_{i}}\right)^{n}_{\otimes}K_{\beta}(w,\vec{x}_{1})
\left(\nabla_{\mu_{j}}\right)^{\ell-n}_{\otimes}K_{\beta}(w,\vec{x}_{2})\prod^{\ell}_{i=1}
g^{\mu_{i}\nu_{i}}(w)\nonumber\\
&&\times
K^{(\ell)}_{\lambda}(w,\vec{x}_{3})_{\nu_{1}\dots\nu_{\ell},\sigma_{1}
\dots\sigma_{\ell}}\prod^{\ell}_{k=1}b_{\sigma_{k}}=\langle
\alpha(\vec{x}_{1})\alpha(\vec{x}_{2})\mathcal{J}^{(\ell)}_{i_{1}\dots
i_{\ell}}\rangle\prod^{\ell}_{k=1}b_{i_{k}}\;,
\end{eqnarray}
where the equality  is based on AdS/CFT correspondence.

On this integral we apply a $d$-dimensional translation
\begin{equation}\label{trans}
    \vec{w}\rightarrow \vec{w}+\vec{x}
\end{equation}
and  inversion (\ref{inv}) to
\begin{equation}\label{inv1}
    w\rightarrow w' ,\quad\vec{x}_{i}\rightarrow
    \vec{x'}_{i},\quad i \in \{1,2\} \;.
\end{equation}
This gives by contracting the $R$'s and by (\ref{norm})
\begin{eqnarray}\label{contr}
   &&\left(x^{2}_{13}x^{2}_{23}\right)^{-\beta}\int
\frac{dw'}{(w'_{0})^{d+1}}\left(\nabla'_{\mu_{i}}\right)^{n}_{\otimes}K_{\beta}(w'-\vec{x'}_{13})
\left(\nabla'_{\mu_{i}}\right)^{\ell-n}_{\otimes}K_{\beta}(w'-\vec{x'}_{23})
(w'_{0})^{\lambda+\ell}\prod^{\ell}_{i=1}b_{\mu_{i}}\;.\quad\quad
   \end{eqnarray}
Since $b_{0}=0$ the $\nabla'$ can be applied to $\vec{x'}_{13}$ ,
$\vec{x'}_{23}$ respectively and it follows
\begin{equation}\label{Ibeta}
   \left(x^{2}_{13}x^{2}_{23}\right)^{-\beta}(-1)^{\ell}
   \langle\vec{b}\cdot\vec{\partial'}_{13}\rangle^{n}
    \langle\vec{b}\cdot\vec{\partial'}_{23}\rangle^{\ell-n}
    I_{\beta,\beta,\lambda+\ell}\left(\vec{x'}_{13},\,\vec{x'}_{23}\right)\;,
\end{equation}
where\footnote{since $\ell$ is even $(-1)^{\ell}=+1$}
\begin{eqnarray}\label{Idelta}
   && I_{\Delta_{1},\Delta_{2},\Delta_{3}}(\vec{x}_{1},\,\vec{x}_{2})=
    \int
\frac{dw}{w_{0}^{d+1}}K_{\Delta_{1}}(w-\vec{x}_{1})
K_{\Delta_{2}}(w-\vec{x}_{2})
w_{0}^{\Delta_{3}}\nonumber\\
    && = \frac{1}{2}\pi^{\mu}\Gamma(\Sigma-\mu)\left[\prod^{3}_{i=1}
    \frac{\Gamma(\Sigma-\Delta_{i})}{\Gamma(\Delta_{i})}\right]
    |\vec{x}_{1}-\vec{x}_{2}|^{-2(\Sigma-\Delta_{3})}
   \end{eqnarray}
and
\begin{equation}\label{dimens}
    \mu=\frac{1}{2}d , \quad
    \Sigma=\frac{1}{2}(\Delta_{1}+\Delta_{2}+\Delta_{3}).
\end{equation}
In our case
\begin{eqnarray}
   &&\Delta_{1}=\Delta_{2}=\beta , \quad \Delta_{3}=\lambda+\ell ,\nonumber\\
    && \Sigma=\beta+\delta+\ell ,\quad \Sigma-\mu=\beta+\ell-1 ,
    \quad (\delta\quad\textnormal{as in (\ref{2pphi})})\\
     && \Sigma-\Delta_{1}=\Sigma-\Delta_{2}=\delta+\ell ,
     \quad\Sigma-\Delta_{3}=\beta-\ell-\delta \;.\nonumber
     \end{eqnarray}
As in (\ref{xi}) we introduce
 \begin{equation}\label{xi1}
    \vec{\xi}=\vec{x'}_{13}-\vec{x'}_{23}=\frac{\vec{x}_{13}}{x_{13}^{2}}-\frac{\vec{x}^{i}_{23}}{x^{2}_{23}}
\end{equation}
and use
\begin{equation}\label{fin}
   \langle\vec{b}\cdot\vec{\partial'}_{13}\rangle^{n}
   \langle\vec{b}\cdot\vec{\partial'}_{23}\rangle^{\ell-n}(\xi^{2})^{-\Delta}=
   (-1)^{n}2^{\ell}(\Delta)_{\ell}(\vec{\xi}^{2})^{-\Delta-\ell}\left\{\langle\vec{b}\cdot\vec{\xi}\rangle^{\ell}
   + \textnormal{trace terms}\right\} \;.
\end{equation}
This yields finally
\begin{eqnarray}
   && g_{\ell}^{-1}\langle
\alpha(\vec{x}_{1})\alpha(\vec{x}_{2})\mathcal{J}^{(\ell)}_{i_{1}\dots
i_{\ell}}\rangle_{AdS}\prod^{\ell}_{k=1}b_{i_{k}}\nonumber \\
    && =\mathcal{N}^{AdS}_{\alpha\alpha\mathcal{J}}
    (x^{2}_{12})^{\delta-\beta}(x^{2}_{13}x^{2}_{23})^{-\delta}\left\{\langle\vec{b}\cdot\vec{\xi}\rangle^{\ell}
   + \textnormal{trace terms}\right\},\label{fin1}\\
     &&
     \mathcal{N}^{AdS}_{\alpha\alpha\mathcal{J}}=2^{\ell}\pi^{\mu}
     \left(\frac{\Gamma(\delta+\ell)}{\Gamma(\beta)}\right)^{2}
     \frac{\Gamma(\beta+\ell-1)\Gamma(\beta-\delta)}{\Gamma(2\delta+2\ell)}
     \frac{1}{2}\sum_{n=0}^{\ell}(-1)^{n}A_{n}\;.\label{finn}
    \end{eqnarray}
The last factor gives using (\ref{Acft}) and omitting the
normalization factor $\mathcal{C}$
\begin{equation}\label{sum}
    \frac{1}{2}\sum_{n=0}^{\ell}(-1)^{n}A_{n}=\frac{1}{2^{\ell+1}(\beta)_{\ell}}\,\,
    {}_{2}F_{1}\left[ {-\ell,-\beta-\ell+1\atop \beta
   } ;\;1
   \right]=\frac{(2\beta+\ell-1)_{\ell}}{2^{\ell+1}(\beta)_{\ell}(\beta)_{\ell}}\;.
\end{equation}
\section{The CFT current $\mathcal{J^{(\ell)}}$}
\quad The current $\mathcal{J^{(\ell)}}$ is traceless symmetric of
rank $\ell$ and has conformal dimension
\begin{equation}\label{dimj}
    \lambda=2\delta+\ell+O(\frac{1}{N})\;.
\end{equation}
For $\ell=2$ this is the energy-momentum tensor of the $O(N)$
conformal sigma model (up to normalization)with exact dimension
$2\delta+\ell$. For $\ell > 2$ (even) the anomalous dimension in
(\ref{dimj}) has been calculated for the interacting theory in
\cite{RuehlLang}. Neglecting the anomalous dimension, the currents
are conserved. To this leading order the currents can be expressed
bilinearly by the $O(N)$ vector field $\phi^{i}(x)$
\begin{equation}\label{jl}
    \mathcal{J^{(\ell)}}(x;a)=\frac{1}{\sqrt{2N}}\sum_{r=0}^{[\ell/2]}
    \sum_{n=0}^{\ell-2r}\mathcal{A}^{(\ell)}_{rn}(a^{2})^{r}\langle\vec{a}
    \cdot\vec{\partial}\rangle^{n}(\vec{\partial})^{r}_{\otimes}\phi^{i}(x)
\langle\vec{a}\cdot\vec{\partial}\rangle^{\ell-n}(\vec{\partial})^{r}_{\otimes}\phi^{i}(x)
\end{equation}
(so that $ \mathcal{J}^{(0)}$ equals $\alpha_{f}$ in
(\ref{alphafree}).  We normalize $\mathcal{A}^{(\ell)}_{rn}$ to
\begin{equation}\label{norm1}
    \mathcal{A}^{(\ell)}_{00}=1
\end{equation}
and get
\begin{equation}\label{solfora}
\mathcal{A}^{(\ell)}_{rn}=\frac{(-1)^{n}\ell!}{2^{r}r!n!(\ell-n-2r)!}
\,\,\frac{(\delta)_{\ell-r}}{(\delta)_{r+n}(\delta)_{\ell-n-r}}\;.
\end{equation}
Its two point function is
\begin{eqnarray}
    &&\langle\mathcal{J^{(\ell)}}(x;a)\,\mathcal{J^{(\ell)}}(0;b)\rangle_{CFT}
    =\mathcal{N}^{CFT}_{\mathcal{J}\mathcal{J}}(x^{2})^{-\lambda}
    \left\{\left[2\frac{\langle\vec{a}\cdot\vec{x}\,
    \rangle\langle\vec{b}\cdot\vec{x}\rangle}{x^{2}}-
    \langle\vec{a}\cdot\vec{b}\rangle\right]^{\ell}\right.\\
   &&\left.\qquad\quad\quad\quad\quad\quad\quad\quad\quad\quad\quad + \textnormal{trace
    terms}\right\}\;.\label{2pf3}\nonumber
\end{eqnarray}
By explicit evaluation of the l.h.s. we obtain
\begin{equation}\label{Njj}
   \mathcal{N}^{CFT}_{\mathcal{J}\mathcal{J}}=2^{\ell}(\ell!)^{2}
   \frac{(2\delta+\ell-1)_{\ell}}{(\delta)_{\ell}}\;.
\end{equation}

\section{The CFT 3-point function of $\alpha$, $\alpha$ and
$\mathcal{J}^{(\ell)}$}\quad In a free conformal $O(N)$ sigma
model the 3-point function can be directly calculated by
contraction of the free fields $\phi^{i}(x)$
\begin{equation}
\langle\alpha(x_{1})\alpha(x_{2})\mathcal{J}^{(\ell)}(x_{3};a)\rangle_{CFT}=
\frac{2^{\ell+2}(\delta)_{\ell}}{\sqrt{2N}}(x^{2}_{12}x^{2}_{13}x^{2}_{23})^{-\delta}
\left\{\langle\vec{\xi}\cdot\vec{a}\rangle^{\ell} + \,\,
\textnormal{traces}\right\}\quad\label{aaj}
\end{equation}
 with $\vec{\xi}$ as
in (\ref{xi}), (\ref{xi1}). In the interacting case we also make
the ansatz
\begin{equation}
   \langle\alpha(x_{1})\alpha(x_{2})\mathcal{J}^{(\ell)}(x_{3};a)\rangle_{CFT}=
\frac{1}{\sqrt{2N}}\mathcal{N}^{CFT}_{\alpha\alpha\mathcal{J}}(x^{2}_{12})^{\delta-\beta}(x^{2}_{13}x^{2}_{23})^{-\delta}
\left\{\langle\vec{\xi}\cdot\vec{a}\rangle^{\ell} + \,\,
\textnormal{traces}\right\}\;.\quad\label{aaj1}
\end{equation}
Contrary to AdS field theory this three-point function is obtained
from a local interaction at second order.

In order to compute the proportionality factor
$\mathcal{N}^{CFT}_{\alpha\alpha\mathcal{J}}$ in (\ref{aaj1}) we
start from the four-point function
\begin{equation}\label{4point}
    \langle\alpha(x_{1})\alpha(x_{2})\phi_{i}(x_{3})\phi_{j}(x_{4})\rangle_{CFT}
    \;,
\end{equation}
which we differentiate with respect to $x_{3}$, $x_{4}$ as
prescribed by (\ref{jl}), then sum over $i=j$ and let finally
$\vec{x}_{3}-\vec{x}_{4}\rightarrow 0$.

There are three graphical contributions \vspace{5mm}
\begin{align}
\begin{picture}(60,30)
\put(-10,10){ \dashline{2}(0,30)(20,10)
\put(20,10){\line(1,1){20}} \put(20,-10){\line(0,1){20}}
\put(20,-10){\line(1,-1){20}} \dashline{2}(20,-10)(0,-30)
\put(-10,-35){$2$} \put(-10,28){$1$} \put(45,-35){$4$}
\put(45,28){$3$} \put(15,-40){$B_{1}$}}
\end{picture}
\qquad \qquad\!\!\!
\begin{picture}(60,30)
\put(-0,10){ \dashline{2}(-10,30)(15,-20)
\put(15,15){\line(2,1){25}} \put(15,-20){\line(0,1){35}}
\put(15,-20){\line(2,-1){25}} \dashline{2}(15,15)(-10,-30)
\put(-20,-35){$2$} \put(-20,28){$1$} \put(45,-35){$4$}
\put(45,28){$3$} \put(10,-40){$B_{2}$}}
\end{picture}
\qquad\qquad
\begin{picture}(60,30)
\put(-0,0){ \dashline{2}(-10,30)(0,20) \dashline{2}(-10,-10)(0,0)
\line(0,1){20} \put(0,0){\line(1,1){10}}
\put(0,20){\line(1,-1){10}} \dashline{2}(10,10)(30,10)
\put(30,10){\line(1,1){20}} \put(30,10){\line(1,-1){20}}
\put(-19,-15){$2$} \put(-19,28){$1$} \put(53,-15){$4$}
\put(53,28){$3$} \put(15,-30){$B_{3}$}}
\end{picture}
\end{align}
\vspace{5mm}

\noindent at leading order $\frac{1}{N}$ to (\ref{4point}). Since
the triangle subgraph of $B_{3}$ vanishes at $d=3$ \cite{LMR,LR}
we neglect it here. $B_{2}$ is obtained by simple crossing of
$B_{1}$: $x_{1}\leftrightarrow x_{2}$.

We define
\begin{equation}\label{uv}
    u=\frac{x^{2}_{12}x^{2}_{34}}{x^{2}_{13}x^{2}_{24}} , \quad
    v=\frac{x^{2}_{14}x^{2}_{23}}{x^{2}_{13}x^{2}_{24}}\;,
\end{equation}
so that in the limit
\begin{eqnarray}
   &&   \vec{x}_{3}-\vec{x}_{4}=\vec{x}_{34} \rightarrow 0 \\
    &&  u\rightarrow 0 , \quad v\rightarrow 1 \;.\label{channel}
\end{eqnarray}
A double power series expansion in $u$ and $1-v$ is
correspondingly appropriate. The crossing $B_{1}\rightarrow B_{2}$
implies
\begin{equation}\label{cross}
   {u\rightarrow \frac{u}{v}\atop
   \;\; v\rightarrow \frac{1}{v}\;.}
   \end{equation}
We remark also that all three graph have been calculated in
another channel in \cite{RuehlLang1}, but we refrain from
performing the nontrivial analytic continuation to our channel
(defined by (\ref{channel})).

The graph $B_{1}$ contributes a constant factor (for the notation
see \cite{LR})
\begin{equation}\label{const}
    \textbf{z}_{1}v(\beta,\delta,\delta)v(1,\beta,2\delta-\beta+1)|_{\beta=2}=2(2\delta-1)
\end{equation}
where $(\frac{\textbf{z}_{1}}{N})^{\frac{1}{2}}$ is the sigma
model coupling constant. In addition $B_{1}$ gives a generalized
hypergeometric function ($\beta=2$)
\begin{equation}\label{hyp}
    (x^{2}_{13}x^{2}_{24})^{-\delta}(x^{2}_{12})^{\delta-\beta}
    \sum_{n=0}^{\infty}\frac{u^{n}}{n!}\,
    \frac{[(\delta)_{n}]^{2}n!}{(2\delta-1)_{2n}(\delta-1)_{n}}
    {}_{2}F_{1}\left[ {\delta+n,\delta+n\atop 2\delta-1+2n} ;\;1-v
    \right]\;.
\end{equation}
For the simple crossing (\ref{cross}) we can use the Euler
identity \footnote{2-term ${}_{2}F_{1}$ relation in \cite{GR}
9.131.1}
\begin{equation}\label{EI}
    v^{-n}{}_{2}F_{1}\left[ {\delta+n,\delta+n\atop 2\delta-1+2n} ;\;1-\frac{1}{v}
    \right]=v^{\delta}{}_{2}F_{1}\left[ {\delta+n,\delta+n-1\atop 2\delta-1+2n} ;\;1-v
    \right]\;.
\end{equation}

For the covariant factor in (\ref{hyp}) we obtain by crossing
\begin{equation}\label{cross2}
(x^{2}_{23}x^{2}_{14})^{-\delta}v^{\delta}=(x^{2}_{13}x^{2}_{24})^{-\delta}\;\;.
\end{equation}
Thus we have simply to add the two Gaussian functions (see
\cite{GR} 9.137.17)
\begin{eqnarray}\label{Gauss}
    &&{}_{2}F_{1}\left[ {\delta+n,\delta+n\atop 2\delta-1+2n} ;\;1-v
    \right]+{}_{2}F_{1}\left[ {\delta+n,\delta+n-1\atop 2\delta-1+2n} ;\;1-v
    \right]\nonumber\\&&=2\,\,{}_{2}F_{1}
    \left[ {\delta+n,\delta+n-1\atop 2\delta-2+2n} ;\;1-v
    \right]\;.
\end{eqnarray}
Since the factor (\ref{const})
\begin{equation}\label{d3}
    2(2\delta-1)|_{d=3}=0
\end{equation}
vanishes, a non vanishing result necessitates a corresponding
pole. Such pole is supplied by the coefficients of
\begin{equation}\label{coef}
    \frac{u^{n}(1-v)^{m}}{n!m!}\;,\nonumber
\end{equation}
whenever
\begin{equation}\label{cof1}
    2n+m\geq 2 \;.
\end{equation}
Next we perform the differentiations
\begin{equation}\label{diff}
    \sum_{\nu=0}^{\ell}\mathcal{A}^{(\ell)}_{0\nu}\,\langle \vec{a}\vec{\partial}_{3}\rangle^{\nu}
    \,\,\langle\vec{a}\vec{\partial}_{4}\rangle^{\ell-\nu}\;.
\end{equation}
The powers $u^{n}$ contain the factor $(x^{2}_{34})^{n}$, which
must be differentiated $2n$ times to be nonzero at $x_{34}=0$.
Doing this we obtain $(a^{2})^{n}$ which contribute to the trace
terms if $n>0$. Thus we may assume $n=0$ if we neglect the trace
terms for the moment.

There remain terms proportional to
\begin{equation}\label{prop}
    (x^{2}_{13}x^{2}_{24})^{-\delta}(1-v)^{m}=
    (x^{2}_{13}x^{2}_{24})^{-\delta-m}(x^{2}_{13}x^{2}_{24}-x^{2}_{14}x^{2}_{23})^{m}\;.
\end{equation}
From (\ref{aaj1}) we know that we must look for the factor
\begin{equation}\label{factor}
    \langle\vec{a}\cdot\vec{\xi}\rangle^{\ell}=\left[\frac{\langle\vec{a}\cdot\vec{x}_{13}\rangle}{x^{2}_{13}}
    -\frac{\langle\vec{a}\cdot\vec{x}_{23}\rangle}{x^{2}_{23}}\right]^{\ell}
    \;.
\end{equation}
For the identification of the normalization factor
$\mathcal{N}^{CFT}_{\alpha\alpha\mathcal{J}}$ it is sufficient to
evaluate only
\begin{equation}
\left[\frac{\langle\vec{a}\cdot\vec{x}_{13}\rangle}{x^{2}_{13}}\right]^{\ell}\;.
\end{equation}
To obtain it we may set
\begin{equation}
    x^{2}_{23}=x^{2}_{24}
\end{equation}
and differentiate only $x_{13},\, x_{14}$ with respect to
$x_{3},\,x_{4}$ before we set also in these terms $x_{3}=x_{4}$.
Thus we have to apply the differentiation (\ref{diff}) instead of
(\ref{prop}) to
\begin{equation}\label{prop1}
    (x^{2}_{23})^{-\delta}(x^{2}_{13})^{-\delta-m}(x^{2}_{13}-x^{2}_{14})^{m}=
    (x^{2}_{23})^{-\delta}\,\sum_{k=0}^{m}
    \binom{m}{k}(x^{2}_{13})^{-\delta-k}(-x^{2}_{14})^{k}\;.
\end{equation}
In the limit $x_{3}=x_{4}$ the result is
\begin{equation}\label{res}
    (x^{2}_{13}\,x^{2}_{23})^{-\delta}2^{\ell}
    \left(\frac{\langle\vec{a}\cdot\vec{x}_{13}\rangle}{x^{2}_{13}}\right)^{\ell}
    \;\sum_{k=0}^{\ell-\nu}(-1)^{k-\ell+\nu}\binom{m}{k}
    \frac{k!}{(k-\ell+\nu)!}(\delta+k)_{\nu}\;.
\end{equation}
Performing the summations (see the Apendix B) we get finally
\begin{equation}\label{N}
    \mathcal{N}^{CFT}_{\alpha\alpha\mathcal{J}}=4(2\delta-1)
    \frac{\ell!(\delta)_{\ell}}{(2\delta-1)_{\ell}}\;.
\end{equation}
The loop of the $\phi^{i}$-fields cancels the perturbative factor
$\frac{1}{N}$.
\section{Renormalization of the higher spin field}
\quad The normalization of the field $h^{(\ell)}$ has been fixed
by the propagators (\ref{2pf3}),\,(\ref{bbl})
\begin{equation}\label{prs}
    \lim_{w_{0}\rightarrow 0} w^{-(\lambda-\ell)}_{0}
    \langle h^{(\ell)}(w)\; h^{(\ell)}(0)\rangle_{AdS}=
    (\mathcal{N}^{CFT}_{\mathcal{J}\mathcal{J}})^{-1}\;
    \langle\mathcal{J}^{(\ell)}(\vec{w})\;
    \mathcal{J}^{(\ell)}(0)\rangle_{CFT}\;\;.
\end{equation}
By AdS/CFT correspondence the CFT 3-point function (\ref{aaj}),
(\ref{aaj1}) where we insert the renormalized current
$\mathcal{J}^{(\ell)}$
\begin{equation}
    \left(\mathcal{N}^{CFT}_{\mathcal{J}\mathcal{J}}\right)^{-\frac{1}{2}}\mathcal{J}^{(\ell)}
\end{equation}
must equal the AdS 3-point function (\ref{fin1}), (\ref{finn})
\begin{equation}
    \frac{1}{\sqrt{2N}}\;\frac{\mathcal{N}^{CFT}_{\alpha\alpha\mathcal{J}}}
    {\left(\mathcal{N}^{CFT}_{\mathcal{J}\mathcal{J}}\right)^{\frac{1}{2}}}
    =g_{\ell}\;\mathcal{N}^{AdS}_{\alpha\alpha\mathcal{J}}\;,
\end{equation}
which implies
\begin{eqnarray}\label{gel}
   &&g_{\ell}=\frac{1}{\sqrt{N}}\;C^{(\ell)}\;\;,\\
   &&C^{(\ell)}=\frac{1}{\sqrt{2}}\frac{\mathcal{N}^{CFT}_{\alpha\alpha\mathcal{J}}}
   {\left(\mathcal{N}^{CFT}_{\mathcal{J}\mathcal{J}}\right)^{\frac{1}{2}}
   \mathcal{N}^{AdS}_{\alpha\alpha\mathcal{J}}}\;\;.
\end{eqnarray}
Since we neglected a graph in the calculation of
$\mathcal{N}^{CFT}_{\alpha\alpha\mathcal{J}}$ which vanishes at
$d=3$, our result makes sense only at $d=3$. We have
\begin{eqnarray}
   && \mathcal{N}^{CFT}_{\alpha\alpha\mathcal{J}}=
   \left\{{2^{\ell+2}(\frac{1}{2})_{\ell}\;\;\textnormal{(free)}\atop
   \;\;\;\;\;\;\;\quad 4\ell(\frac{1}{2})_{\ell}\;\;\;\;\;\;\textnormal{(interacting)\;\;,}}\right.\\
    && \mathcal{N}^{CFT}_{\mathcal{J}\mathcal{J}}=\;\;\;2^{3\ell-1}(\ell!)^{2}\;\;,\\
     && \mathcal{N}^{AdS}_{\alpha\alpha\mathcal{J}}=
     \left\{\frac{\pi^{2}}{2\ell}\left(\frac{\Gamma(\ell+\frac{1}{2})}{\ell!}\right)^{2}
     \;\;\textnormal{(free)}\atop \;\;\;\;\frac{\pi^{2}}{2(\ell+2)}\,\,
     \left(\frac{\Gamma(\ell+\frac{1}{2})}{(\ell+1)!}\right)^{2}
     \;\;\textnormal{(interacting)\;\;,}\right.
    \end{eqnarray}
implying
\begin{equation}\label{clf}
    C^{(\ell)}=\left\{{\frac{2^{-\frac{1}{2}\ell+2}}{\pi^{3}}\;\ell\;
    \frac{\ell!}{(\frac{1}{2})_{\ell}}\quad\quad
    \textnormal{(free)}\atop\frac{4\sqrt{2}}
    {\pi^{3}}\;\ell(\ell+1)^{2}(\ell+2)\;\frac{\ell!}{(\frac{1}{2})_{\ell}}\quad
\textnormal{(interacting)}\;.}\right.
\end{equation}
Finally we renormalize
\begin{equation}
H^{(\ell)}(z)=C^{(\ell)}h^{(\ell)}(z)\;\;.
\end{equation}

\subsection*{Acknowledgements}
This work is supported in part by the German Volkswagenstiftung.
The work of R.~M. was supported by DFG (Deutsche
Forschungsgemeinschaft) and in part by the INTAS grant
\#03-51-6346 .

\section*{Appendix A\\Spin 4 current in $AdS_{4}$ in details}
\setcounter{equation}{0}
\renewcommand{\theequation}{A.\arabic{equation}}

We will define the traceless fourth rank tensor constructed from
four dimensional on-shell scalar field $\sigma(z^{\mu})$ in the
following way
\begin{eqnarray}
  T_{\mu\nu\lambda\rho}^{\textnormal{traceless}} &=&
  T_{\mu\nu\lambda\rho}-\frac{3}{8}\left(g_{\mu(\nu}T_{\lambda\rho)}+T_{\mu(\nu}g_{\lambda\rho)}\right)
  +\frac{1}{16}g_{\mu(\nu}g_{\lambda\rho)}T\;\;, \label{traceA}\\
  T_{\mu\nu} &=& T^{\alpha}_{\alpha\mu\nu}\quad , \quad
  T=T^{\mu\nu}_{\mu\nu}\nonumber\;\;.
\end{eqnarray}
The conservation law which we will check below is
\begin{equation}\label{cl}
\nabla^{\mu}T_{\mu\nu\lambda\rho}^{\textnormal{traceless}} =
  \nabla^{\mu}T_{\mu\nu\lambda\rho}-\frac{3}{8}\left(\nabla_{(\nu}T_{\lambda\rho)}+\nabla^{\mu}T_{\mu(\nu}g_{\lambda\rho)}\right)
  +\frac{1}{16}g_{(\nu\lambda}\nabla_{\rho)}T=0\;\;.
\end{equation}
Finally we list here the most important on-shell relations (some
of them are due to $\Box \sigma(z)=\frac{2}{L^{2}}\sigma(z)$) we
will use
\begin{eqnarray}
  &&\left[\Box ,
\nabla_{\mu}\right]\sigma(z)=\frac{3}{L^{2}}\nabla_{\mu}\sigma(z) \label{com1}\;, \\
  &&\left[\nabla_{\mu},\nabla^{3}_{(\nu\lambda\rho)}\right]\sigma(z)=
  \frac{3}{L^{2}}g_{\mu(\nu}\nabla^{2}_{\lambda\rho)}\sigma(z)
  -\frac{3}{L^{2}}g_{(\nu\lambda}\nabla_{\rho)}\nabla_{\mu}\sigma(z)\;,\\
  &&\nabla^{3}_{(\mu\lambda\rho)}\sigma(z)=\nabla^{2}_{(\lambda\rho)}\nabla_{\mu}\sigma
  + \frac{1}{3L^{2}}g_{\mu(\rho}\nabla_{\lambda)}\sigma(z) -
  \frac{1}{L^{2}}g_{\lambda\rho}\nabla_{\mu}\sigma(z)\;,\\
  &&\nabla^{\mu}\nabla^{2}_{(\mu\lambda\rho)}\sigma(z)=
  \frac{28}{3L^{2}}\nabla^{2}_{(\lambda\rho)}\sigma(z)
  - \frac{8}{3l^{4}}g_{\lambda\rho}\sigma(z)\;,\\
  &&\left[\nabla_{\mu},\nabla^{3}_{(\nu\lambda\rho)}\right]\nabla^{\mu}\sigma(z)=
  \frac{12}{L^{2}}\nabla^{3}_{(\nu\lambda\rho)}\sigma(z)
  -\frac{9}{L^{4}}g_{(\nu\lambda}\nabla_{\rho)}\sigma(z)\;,\\
  &&g^{\lambda\rho}\nabla^{3}_{(\mu\lambda\rho)}\sigma(z)=\frac{4}{L^{2}}
  \nabla_{\rho}\sigma(z)\;,\\
  &&g^{\lambda\rho}\nabla^{4}_{(\mu\nu\lambda\rho)}\sigma(z)=
  \frac{20}{3L^2}\nabla^{2}_{(\mu\nu)}\sigma(z)-\frac{4}{3L^{4}}\sigma(z)\;. \label{trace1}
\end{eqnarray}

Now we can construct directly the conserved spin 4 traceless
current. First of all we note that from four derivatives we can
construct only three bilinear combinations
\begin{eqnarray}
  T^{0,4}_{\mu\nu\lambda\rho} &=&
  \sigma\nabla_{(\mu}\nabla_{\nu}\nabla_{\lambda}\nabla_{\rho)}\sigma\;,\label{04}\\
  T^{1,3}_{\mu\nu\lambda\rho}&=&
  \nabla_{(\mu}\sigma\nabla_{\nu}\nabla_{\lambda}\nabla_{\rho)}\sigma\;,\label{13}\\
  T^{2,2}_{\mu\nu\lambda\rho}&=&
  \nabla_{(\mu}\nabla_{\nu}\sigma\nabla_{\lambda}\nabla_{\rho)}\sigma\;.\label{22}
\end{eqnarray}
For constructing the  conserved (on-shell) combination of the
traceless parts of these tensors we need first of all the on-shell
value of their first and second traces
\begin{eqnarray}
 &&T^{0,4}_{\lambda\rho} =\frac{20}{3l^2}\sigma(z)\nabla^{2}_{(\mu\nu)}\sigma(z)-
 \frac{4}{3L^{4}}\sigma^{2}(z)\quad , \quad T^{0,4}=\frac{8}{L^{4}}\sigma^{2}(z)\;,\\
 &&T^{1,3}_{\lambda\rho}=\frac{1}{2}\nabla^{\mu}\sigma\nabla^{2}_{(\lambda\rho)}
 \nabla_{\mu}\sigma
 +\frac{13}{6L^{2}}\nabla_{\lambda}\sigma\nabla_{\rho}\sigma - \frac{1}{6L^{2}}g_{\lambda\rho}
 \sigma\nabla_{\mu}\sigma ,\quad T^{1,3}=\frac{4}{L^{2}}\nabla^{\mu}\sigma
 \nabla_{\mu}\sigma\;,\quad\quad\quad\\
 &&T^{2,2}_{\lambda\rho} = \frac{2}{3}\nabla_{\lambda}\nabla^{\mu}\sigma\nabla_{\rho}
 \nabla_{\mu}\sigma +
 \frac{2}{3L^{2}}\sigma\nabla^{2}_{(\lambda\rho)}\sigma , \quad
  T^{2,2}=\frac{2}{3}\nabla^{2(\mu\nu)}
  \sigma\nabla_{(\mu\nu)}^{2}\sigma+\frac{4}{3L^{4}}\sigma^{2}\;.
\end{eqnarray}
Then introducing the following third rank symmetric tensor
bilinear terms
\begin{eqnarray}
  \textbf{A} &=& \nabla_{(\nu}\nabla^{\mu}\sigma\nabla^{2}_{\lambda\rho)}\nabla_{\mu}\sigma ,
  \quad \textbf{a}=\nabla_{(\nu}\sigma\nabla^{2}_{\lambda\rho)}\sigma\;,\\
 \textbf{B}&=& g_{(\nu\lambda}\nabla_{\rho)}\left(\nabla^{2(\mu\nu)}\sigma\nabla^{2}_
 {(\mu\nu)}\sigma\right) , \quad \textbf{b}=g_{(\nu\lambda}\nabla_{\rho)}
 \left(\nabla^{\mu}\sigma\nabla_{\mu}\sigma\right)\;,\\
 \textbf{C}&=&\nabla^{\mu}\sigma\nabla^{3}_{(\nu\lambda\rho)}\nabla_{\mu}\sigma , \quad
 \textbf{c}= \sigma\nabla^{3}_{(\nu\lambda\rho)}\sigma ,\quad
 \textbf{d}=g_{(\nu\lambda}\nabla_{\rho)}\left(\sigma^{2}\right)
\end{eqnarray}
and using (\ref{traceA})-(\ref{trace1}), we obtain the following
on-shell relations
\begin{eqnarray}
  &&\nabla^{\mu}T^{2,2\textnormal{traceless}}_{\mu\nu\lambda\rho}=
  \frac{1}{2}\textbf{A}-\frac{1}{12}\textbf{B}+\frac{23}{4L^{2}}\textbf{a}
  -\frac{9}{8L^{2}}\textbf{b}-\frac{1}{4L^{2}}\textbf{c}-\frac{19}{24L^{4}}
  \textbf{d}\;, \label{n22}\\
  &&\nabla^{\mu}T^{1,3\textnormal{traceless}}_{\mu\nu\lambda\rho}=\frac{9}{16}\textbf{A}-
  \frac{3}{32}\textbf{B}+ \frac{1}{16}\textbf{C}+\frac{51}{8L^{2}}\textbf{a}
  -\frac{11}{8L^{2}}\textbf{b}+\frac{1}{2L^{2}}\textbf{c}-\frac{13}{8L^{4}}
  \textbf{d}\;,\quad\quad\quad\label{n13}\\
&&\nabla^{\mu}T^{0,4\textnormal{traceless}}_{\mu\nu\lambda\rho}=\textbf{C}
-\frac{3}{2L^{2}}\textbf{a}
  -\frac{7}{4L^{2}}\textbf{b}+\frac{25}{2L^{2}}
  \textbf{c}-\frac{47}{4L^{4}}\textbf{d}\;.\label{n04}
\end{eqnarray}
Now we can see that the following unique combination of
(\ref{04})-(\ref{22}) is conserved
\begin{eqnarray}\label{s4final}
&&T^{s=4,\textnormal{traceless}}_{\mu\nu\lambda\rho}=
T^{2,2\textnormal{traceless}}_{\mu\nu\lambda\rho}
-\frac{8}{9}T^{1,3\textnormal{traceless}}_{\mu\nu\lambda\rho}
+\frac{1}{18}T^{0,4\textnormal{traceless}}_{\mu\nu\lambda\rho}\;,\\
&&\nabla^{\mu}T^{s=4,\textnormal{traceless}}_{\mu\nu\lambda\rho}=0\;.
\label{ccond}
\end{eqnarray}
The expression (\ref{s4final}) for the current is again in
agreement with the flat space case general formula after a
replacement of ordinary derivatives by covariant ones (compare the
coefficients in (\ref{s4final}) with the solution (\ref{4d}) and
overall factor $\frac{1}{36}$).

\section*{Appendix B}
\setcounter{equation}{0}
\renewcommand{\theequation}{B.\arabic{equation}}
We have to perform the triple summation
\begin{equation}\label{sumation}
    2(2\delta-1)\sum_{\nu=0}^{\ell}\binom{\ell}{\nu}
    \frac{(\delta)_{\ell}}{(\delta)_{\nu}(\delta)_{\ell-\nu}}
    \sum_{m=0}^{\ell}\frac{(\delta)_{m}(\delta)_{m-1}}{(2\delta-1)_{m-1}m!}
    \sum_{k=\ell-\nu}^{m}(-1)^{k}\frac{k!}{(k-\ell+\nu)!}\binom{m}{k}(\delta+k)_{\nu}\;,
\end{equation}
where the first term arises from $\mathcal{A}^{(\ell)}_{0\nu}$ in
(\ref{diff}), the second term from the Gaussian hypergeometric
series (\ref{Gauss}) and the last one from the expansion
(\ref{res}).

The $k$-summation is with
\begin{eqnarray}
   && \quad\quad\quad\quad\quad k=\ell-\nu+r\nonumber \\
    && \sum_{r=0}^{m+\nu-\ell}\frac{\Gamma(-m+\ell-\nu+r)
    \Gamma(\delta+\ell+r)}{\Gamma(-m)\Gamma(\delta+\ell-\nu+r)r!}=
    \nu!m!\frac{(\delta)_{\ell}}{(\delta)_{m}}\,\frac{(-1)^{m}}{(\ell-m)!(\nu+m-\ell)!}.
    \quad\quad\quad
   \end{eqnarray}
The $m$-summation is
\begin{equation}
    \nu!(\delta)_{\ell}\sum_{m=\ell-\nu}^{\ell}\frac{(-1)^{m}}{(\ell-m)!(\nu+m-\ell)!}
    \,\frac{(\delta)_{m-1}}{(2\delta-1)_{m-1}}.
\end{equation}
After substitution of
\begin{equation}
    m=\ell-\nu+s \;,
\end{equation}
it follows
\begin{eqnarray}
   &&= 2 (-1)^{\ell-\nu} \frac{(\delta)_{\ell}(\delta-1)_{\ell-\nu}}{(2\delta-2)_{\ell-\nu}}
   \sum_{s=0}^{\nu}\frac{(-\nu)_{s}(\delta-1+\ell-\nu)_{s}}{s!(2\delta-2+\ell-\nu)_{s}} \\
    &&= 2 (-1)^{\ell-\nu}\frac{(\delta)_{\ell}(\delta-1)_{\ell-\nu}
    (\delta-1)_{\nu}}{(2\delta-2)_{\ell}}\;.
\end{eqnarray}

The $\nu$-summation is
\begin{eqnarray}
   && 2(\delta)_{\ell}\sum_{\nu=0}^{\ell}\binom{\ell}{\nu}
   \frac{(\delta)_{\ell}}{(\delta)_{\nu}(\delta)_{\ell-\nu}}\;(-1)^{\ell-\nu}\;
   \frac{(\delta-1)_{\ell-\nu}(\delta-1)_{\nu}}{(2\delta-2)_{\ell}} \\
    && = 2 \frac{[(\delta)_{\ell}]^{2}}{(2\delta-2)_{\ell}}\sum_{\nu=0}^{\ell}
    \frac{(-\ell)_{\nu}}{\nu!}\frac{(\delta-1)_{\nu}}{(\delta)_{\nu}}
    \frac{(\delta-1)_{\ell-\nu}}{(\delta)_{\ell-\nu}}\;.
   \end{eqnarray}
This sum can be decomposed into two ${}_{2}F_{1}(1)$ sums by
\begin{equation}
    \frac{(\delta-1)_{\nu}}{(\delta)_{\nu}}\;
    \frac{(\delta-1)_{\ell-\nu}}{(\delta)_{\ell-\nu}}=\frac{\delta-1}{2\delta+\ell-2}
    \left[\frac{(\delta-1)_{\nu}}{(\delta)_{\nu}} +
    \frac{(\delta-1)_{\ell-\nu}}{(\delta)_{\ell-\nu}}\right]
\end{equation}
and we use for even $\ell$
\begin{equation}
    \frac{(-\ell)_{\nu}}{\nu!}=\frac{(-\ell)_{\ell-\nu}}{(\ell-\nu)!}
\end{equation}
to obtain
\begin{equation}
    2\;\frac{\ell!(\delta)_{\ell}}{(2\delta-1)_{\ell}}\;,
\end{equation}
which leads to (\ref{N}) immediately.

\end{document}